\documentclass[12pt]{article}
\title{The Parametric Solution of Underdetermined linear ODEs}
\author{Thomas Wolf\\ Department of Mathematics,
Brock University\\ 500 Glenridge Avenue, St.Catharines,\\
Ontario, Canada L2S 3A1\\
email: twolf@brocku.ca}

\begin{document}
\pagestyle{empty}
\maketitle
\begin{abstract}

The purpose of this paper is twofold. An immediate practical use of
the presented algorithm is its applicability to the parametric
solution of underdetermined linear ordinary differential equations
(ODEs) with coefficients that are arbitrary analytic functions in the
independent variable. A second conceptual aim is to present an
algorithm that is in some sense dual to the fundamental Euclids
algorithm, and thus an alternative to the special case of a
Gr\"{o}bner basis algorithm as it is used for solving linear ODE-systems. 
In the paper Euclids algorithm and the new `dual version' are
compared and their complementary strengths are analysed
on the task of solving underdetermined ODEs. An
implementation of the described algorithm is interactively accessible
under \cite{Wuode}.



\end{abstract}


\section{Introduction} \label{intro}
Underdetermined ordinary and partial differential equations
(ODEs/PDEs) are typical objects of investigation in control theory
but also overdetermined systems of
equations where originally the number of equations is larger than the
number of functions may turn into underdetermined problems in the 
course of their solution. Examples are the conditions for
Lie-symmetries and conservation laws of linearizable differential
equations. These overdetermined systems of conditions have solutions
that involve arbitrary functions of one or more variables, i.e.\ in
the process of solving these systems towards its conclusion,
underdetermined problems often occur. 

Algorithms for solving underdetermined linear equations and systems
(ODEs, PDEs, multidimensional discrete,..) are known
(\cite{CQR05},\cite{PQ04},\cite{FO98}). There also exists efficient
software by Daniel Robertz et al.\ \cite{CQRwww}, \cite{CQR07}. The
purpose of this paper is to describe an alternative algorithm that is
elegant, efficient and in some sense complementary to the
fundamental Euclids algorithm which is the basis for the Gr\"{o}bner
basis algorithms that are used in other implementations.
An open question is whether the new algorithm can be generalized to partial
differential equations (PDEs) and thus not only be a complimentary
algorithm for the non-commutative differential Gr\"{o}bner basis
algorithm for solving ODEs but also for solving PDEs.

An implementation of algorithms described in this paper is accessible
online under \cite{Wuode} and is part of the package {\sc Crack}
(\cite{crack}) for solving overdetermined systems. 
As we will describe, a new feature of
these programs are flags which allow to prevent denominators or that
allow to reduce the number of terms in the solution. This can be
achieved by absorbing explicit $x$-dependent factors into new
functions that are introduced during the computation.

We consider a single linear ODE
\begin{eqnarray}
	 0   & = & \sum_i^r D_i f_i \, + \, c_{00}(x), \ \ \ \ r > 1,  \label{uode} \\
        D_i  & = & \sum_{j=0}^{n_i}c_{ij}(x)\left(\frac{d}{dx}\right)^j, \ \ \ \ n_i > 0
        \nonumber 
\end{eqnarray}
for functions $f_1 \ldots f_r$ of the independent variable $x$.
The equation may be homogeneous $(c_{00} = 0)$ or
inhomogeneous $(c_{00} \neq 0)$. With the requirement that the ODE 
is underdetermined we only assume that at least two functions $f_i$
are involved, i.e.\ $r>1$. The differential order of each $f_i$ is
$n_i$. For simplicity we assume that all coefficients $c_{ij}$ are
sufficiently often differentiable, at least $\sum_i n_i$ times.
An upper index in round brackets indicates the number of 
differentiations, i.e.\ $\frac{d^nf_i}{dx^n} = f_i^{(n)}$ and
low order derivatives are denoted by apostrophe, e.g.\ 
$\frac{d^2f_i}{dx^2} = f_i''$.

The task is to find the general solution of the ODE (\ref{uode})
in the form of explicit differential expressions $F_i$
\begin{equation}
f_i = F_i(x,h_1(x),\ldots,h_{r-1}(x)), \ \ i=1,\ldots,r \label{usol}
\end{equation}
in terms of parametric functions $h_1(x),\ldots,h_{r-1}(x)$ which are either
all free or one of them, say $h_1(x)$, having to satisfy an ODE and 
the other $h_j(x)$ being free. 

For example, the general solution of the ODE
\[0 = f''x^2  + g''x - g'x^2  + f + 3x\]
for $f=f(x), \ g=g(x)$ can be written in the form
\begin{eqnarray}                                 
f&=&\frac{x}{x^8 - 2x^6 + 7x^4  - 6x^2  + 9}
    \left((x^5 - x^3 + 3x)h'' - (x^6 + x^4 + 3x^2 - 6)h' \right.  \\ 
 & &\left. + (3x^5 + 3x^3 + 17x)h - 3x^8 + 3x^6 - 16x^4  + 9x^2\right) ,\nonumber \\
g&=&\frac{x}{2(x^8 - 2x^6 + 7x^4 - 6x^2 + 9)}
    \left((- 2x^6 + 2x^4 - 6x^2)h'' \right.  \\ 
 & &\left. + (8x^5 - 4x^3 )h' - (14x^4 + 14x^2 + 6)h 
    + 4x^7 + x^5 + 3x^3 - 27x\right) \label{ex1}
\end{eqnarray}
where $h=h(x)$ is an arbitrary function. The form of the solution is
not unique.  Because $h(x)$ is arbitrary, replacing $h$ by a
differential expression in one or more arbitrary functions would give
a solution too, but a solution with expressions of higher differential
order in more arbitrary functions without the solution being more
general. One naturally seeks a general solution which involves only as
few as possible arbitrary functions and lowest order derivatives of
them. But even this requirement does not give a unique solution. One
might want to minimize the highest order of all parametric functions
or, for example, the sum of all orders of all parametric functions.

Furthermore, the occurring function(s) could be scaled to modify the form
of the solution, for example, to make it denominator free. In the
above case replacing $h(x)$ by $(x^8 - 2x^6 + 7x^4 - 6x^2 + 9)^3p(x)$
with an arbitrary function $p(x)$ would make the solution polynomial
but increase its size, i.e.\ the total number of terms in the coefficients
of the parametric functions on the right hand sides.

In the following section the algorithm is given, followed by comments
on its characteristics. In section \ref{vector} and \ref{euclid} we
formulate the new algorithm and Euclids algorithm in vector notation
in order to compare them in detail in section (\ref{compare}). We show
that both are essentially different but also that they compliment each
other in the sense that one can give a criterion under which
circumstances which of both is better suited.  Another option is to
combine both algorithms in a hybrid version.

Finally, in section (\ref{application}) an application is described
which arose from a classification of hyperbolic evolutionary vector
PDEs.

\section{The Algorithm} \label{algorithm}
\subsection{Outline}
The essence of the algorithm is 
\begin{itemize}
\item to partition the ODE into a total derivative and an algebraic remainder, 
\item to introduce a new function $f_{r+1}$ such that the total
      derivative part is $df_{r+1}/dx$ and thus to write the ODE as a
      system of 2 equations: one equation defining  $f_{r+1}$ and one
      re-formulating the ODE in terms of $f_{r+1}$,
\item to use the re-formulated ODE to eliminate and substitute another
      function and thus to arrive again at a single ODE for the same
      number of functions which in some sense is closer to be solvable
      algebraically than the ODE before.

\end{itemize}
These steps are repeated until either the ODE involves only one
function and thus is not underdetermined anymore, or until one
function occurs purely algebraically and thus allowing the ODE to be
solved for that function. The following is a more detailed pseudo code description.

\subsection{Pseudocode}
\begin{tabbing}
{\bf Input}\ \= $\bullet$ list of functions $f_1,\ldots,f_r$ of $x$, \\
             \> $\bullet$ linear ODE $0=\omega$ in $f_i$ (like (\ref{uode})) \\
{\bf Body}   \> \\
\ \ \= $L:=\{ \}$  \ \ \= \% $L$ will be a list of substitutions \\
    \> $s:=r$          \>  \\
    \> {\bf while} \= (the ODE $0=\omega$ involves at least two $f_i$) {\bf and} \\
    \>             \> (differential orders $n_i > 0 \ \forall f_i$) {\bf do} \\
    \> \ \ \= $\bullet$ \= Factor out $\frac{d}{dx}$ from $\omega$ once as far as possible \\
    \>     \>           \> by introducing a new function $f_{s+1}(x)$ and \\
    \>     \>           \> by computing expressions $b_i$:
\end{tabbing}
\begin{equation}
    0 = \omega = f_{s+1}^{\ \ \ '} + \sum_{i=1}^sf_ib_i + a_{00}            \label{e2}
\end{equation}
\ \ \ \ \ \ where $f_{s+1}, b_i$ are given through: 
\begin{eqnarray}
    f_{s+1} & = & \sum_{i=1}^s\sum_{j=0}^{n_i-1}f_i^{(n_i-1-j)}\sum_{k=0}^j
		 (-1)^{(j+k)}a_{i,n_i-k}^{(j-k)}                    \label{e3}  \\
    b_i & = & \sum_{k=0}^{n_i}(-1)^{(n_i+k)}a_{i,n_i-k}^{(n_i-k)}   \label{e4}
\end{eqnarray}
\begin{tabbing}
\ \ \= \ \ \= $\bullet$ \= {\bf if} $b_i=0$ for all $i$ {\bf then} \\
    \>     \>    \> \ \ \=the ODE \ $0=\omega$ is exact (apart from $a_{00}$), i.e.\ \\
    \>     \>    \>     \> consider new ODE $0 = \hat{\omega} := f_{s+1} + \int a_{00}\ dx$ \\
    \>     \>    \>     \> where $f_{s+1}$ is only an abbreviation defined in (\ref{e3}) \\
    \>     \>    \> {\bf else} \\
    \>     \>    \> \ \ $\circ$ \= regard $f_{s+1}$ as a new unknown function, \\
    \>     \>    \> \ \ $\circ$ \> solve (\ref{e2}) for one function $f_j$ of $f_1,..,f_s$ that\\
    \>     \>    \>   \> $-$ has a non-vanishing coefficient $b_i$ in (\ref{e2}), and \\
    \>     \>    \>   \> $-$ is of lowest possible order in $\omega$ and thus in (\ref{e3}), 
\end{tabbing}
\begin{equation}
    f_j = - \frac{1}{b_j}\left(f_{s+1}^{\ \ \ '} + \sum_{i\neq j}f_ib_i + a_{00}\right)
    \label{e5}
\end{equation}
\begin{tabbing}
\ \ \= \ \ \= $\bullet$ \=  \ \ $\bullet$ \= \kill 
    \>     \>    \>  \ \ $\circ$ \> use (\ref{e5}) to substitute $f_j$ in 
                         (\ref{e3}) to get a new ODE \\
    \>     \>    \>   \> $0 = \hat{\omega} := - f_{s+1} + \sum_{i=1}^s\sum_{j=0}^{n_i-1} \ldots$ \\
    \>     \>    \>   \> for functions $f_1,..,f_{j-1},f_{j+1},..,f_s,f_{s+1}$ \\
    \>     \>    \>  \ \ $\circ$ \> update \ \  
                         $\omega:=\hat{\omega}, \ \ s:=s+1, \ \ L:=\{f_j=..\} \cup L$ \\
    \> {\bf end} \\
    \> {\bf if} the ODE involves a function $f_j$ purely algebraically {\bf then}\\
    \> solve for $f_j$ and add it to $L:\ \ \ L:=\{f_j=..\} \cup L$ \\
    \>  \\
    \> $h(x) := f_s(x)$ \ \ \= \% to remember the last introduced function \\
    \> $P := L$             \> \% to remember the complete list of substitutions $L$ \\
    \> \% next substitutions each as stored in the first \\
    \> \% element of $P$ are performed in the rest of $P$ \\
    \> {\bf while} $s > r$ {\bf do} \\
    \> \ \ \= $P := rest(P)|_{f_s=\ldots \ \ {\rm \% \ as\ given\ in} \ first(P)} $ \\
    \>     \> $s := s-1$ \\
    \> {\bf end} \\
{\bf end} \ \ \% of Body \\ 
\\
{\bf Output} \ \= $\bullet$ \= $l$ \ \ \   \= \% list of new functions \\
               \> $\bullet$ \> $L$         \> \% the complete list of substitutions \\
               \> $\bullet$ \> $P$         \> \% All initial functions $f_1,..,f_r$ are either \\
               \>           \>             \> \% parametric, or are given in $P$ in terms of $h$ \\
               \> $\bullet$ \> $0=\omega(x,h(x))$ \= \% only if the first while loop \\
               \>           \>                    \> \% terminates due to $\omega$ not \\
               \>           \>                    \> \% being underdetermined anymore
\end{tabbing}

\subsection{Comments}  \label{comments}
Before comparing the algorithm with Euclids algorithm a few comments are 
necessary. 
\begin{itemize}
\item All steps involve only algebra or differentiations with only one exception: 
  if the homogeneous part of the ODE is exact then the integral of the
  inhomogeneous part is taken. But this integral does not have to be evaluated, i.e.\
  to be expressed in terms of elementary functions. It can stay in a symbolic 
  unevaluated form for the algorithm to continue.
\item The coefficients $a_{ij}$ can be arbitrary explicit functions of $x$,
  and involve, for example, $\sin$ or $\log$ with the only condition that
  it must be decidable whether expressions involving these functions and their
  derivatives are zero or not.
\item The transformation of the ODE in each step of the first while loop
  is reversible, i.e.\ the new ODE is equivalent to the previous one. Therefore the
  obtained solution is the general one.
\item The algorithm terminates.\\
  Proof: \\
  We consider how the total sum $\sum_{i=1}^s n_i$ of differential 
  orders $n_i$ of all functions $f_i$ changes during execution. 
  After substitution of $f_j$ with (\ref{e5}) in (\ref{e3}) the new
  differential orders $\hat{n}_i$ of functions $f_i$ that occur 
  in the new ODE are:
  \[ \hat{n}_i \; \; \; \left\{ \begin{array}{lll}
             = & n_{j}                             & \mbox{for $i = s+1$} \\
	  \leq & \max(n_j-1, n_i - 1) =  n_i - 1   & \mbox{for $i \leq s, \; b_i \neq 0$} \\
	     = & n_i - 1                           & \mbox{for $i \leq s, \; b_i = 0$} .
	                        \end{array}
  \right. \]
  The function $f_j$ of order $n_j$ gets replaced by a new function
  $f_{s+1}$ which then occurs with same order $n_j$, but the order of
  all other functions is lowered by at least one because we choose $j$
  such that $n_{j} = \min(n_k) \; ( \forall k$ with $b_k \neq 0)$,
  i.e.\ we have $\hat{n}_i \leq n_i - 1$ also for the $f_i$ with $b_i
  \neq 0$. Because the ODE has at least two functions, the total sum
  of derivatives is decreasing. The algorithm is therefore finite.
\item The algorithm results in an ODE for a single function iff the differential
  operators $D_i$ in (\ref{uode}) have a common differential factor. The remaining
  ODE is of the same order as the common factor, i.e.\ its order can not be higher
  than the order of the original ODE (\ref{uode}).


\item The algorithm naturally splits into two
  parts, a part A) establishing a list $L$ of substitutions (\ref{e5})
  in the first while loop and part B) performing the substitutions in
  the second while loop to obtain an explicit solution. The
  first part is executed very fast (see next section for more
  details), the second may take longer for higher order ODEs because
  expressions typically grow with each substitution, often
  exponentially. An example is given in the appendix.
 
For many applications the list $L$ of substitutions may even be of
higher practical value than the explicit formulas $f_i =
F_i(x,h_1(x),\ldots,h_{r-1}(x)), \ \ i=1,\ldots,r $ resulting from B).
In general, the list $L$ is a much shorter representation of the
parametric solution of the ODE than the explicit solution itself and
thus is more useful as a solution, like in the following scenario.

Let us assume that we have a large algebraic expression in terms of
the original functions $f_1,..,f_r$ that has to be evaluated modulo
the underdetermined ODE (\ref{uode}).  Instead of replacing the $f_i$
directly by their large explicit expressions as given in the list $P$
it is usually much better to perform successively the substitutions
stored in $L$ in the order they were derived, allowing cancellations
to happen after each individual substitution. Also, if numerical
computations are to be done, it is much faster to compute the sequence
of substitutions than the explicit expressions.

\end{itemize}

\section{A vector representation} \label{vector}
To perform the iteration process efficiently, i.e.\ to replace
functions by algebraic combinations of other functions (\ref{e5}) 
in the ODE (\ref{e3})
without needing to perform any differentiations, the ODE has to be
represented in a form where $D:=\frac{d}{dx}$ is factored out as far
as possible. As we will see further down, this is also the appropriate 
representation for using the Euclidean algorithm to solve the 
underdetermined ODE. We write the ODE in the form
\begin{eqnarray}
	 0   & = & \sum_i^r A_i f_i \, + \, a_{00}(x)  \label{d1} \\
        A_i  & = & D \tilde{A}_i + a_{i0}(x)           \label{d2} \\
 \tilde{A_i} & = & D^{n_i-1} a_{in_i}(x)+\ldots+Da_{i2}(x)+a_{i1}(x)   \label{d3} 
\end{eqnarray}
Replacing, for example, $f_1=w(x)\hat{f}_1$ would only require multiplications
$\hat{a}_{1k}:=wa_{1k}$ to update this representation and no differentiations.

In this notation a single step in the first while loop of the new method consists of
\begin{itemize}
\item introducing a new function $f_{r+1}$ through 
  \begin{equation}
  f_{r+1}=\sum_i \tilde{A}_i f_i    \label{d4} 
  \end{equation}
  giving the ODE the form
  \begin{equation}
  0 = D f_{r+1} + \sum_i a_{i0}(x)f_i + a_{00}.    \label{d5}
  \end{equation}
\item regarding the defining relation (\ref{d4}) for the new function
  as the new ODE and using the old ODE (\ref{d5}) for a substitution
  of an `old' function. For that we choose one $a_{j0}$ of the non-vanishing
  $a_{i0}$ in (\ref{d2}) for which the order $n_j$ of the corresponding
  $f_j$ is minimal. Thus, performing the substitution 
  \begin{equation}
  f_j = -\frac{1}{a_{j0}}\left(Df_{r+1} + \sum_{i \neq j}a_{i0}f_i + a_{00}\right)
  \label{d6}
  \end{equation}
  in the definition (\ref{d4}) gives the new ODE
  \begin{equation}
  0 = - f_{r+1} + \sum_{i \neq j} \tilde{A}_i f_i
                - \sum_{i \neq j} \tilde{A}_j \left(\frac{a_{i0}}{a_{j0}}f_i\right)
		-		  \tilde{A}_j \left(\frac{    1 }{a_{j0}}Df_{r+1}\right)
		-		  \tilde{A}_j \left(\frac{a_{00}}{a_{j0}}\right) .
  \label{d7}
  \end{equation}

\end{itemize}
In vector notation the above two steps change the differential operators 
(assuming for simplicity of notation $j=1$) according to:
  \begin{equation}
    \left(\begin{array}{c} A_1 \vspace{12pt} \\ A_2 \\ : \\ A_r \end{array}\right) \rightarrow 
    \left(\begin{array}{c} A_{r+1}\vspace{12pt} \\ A_2 \\ : \\ A_r \end{array} \right) = 
    \left(\begin{array}{c} \;\;-1-\tilde{A}_1\frac{1}{a_{10}}D\vspace{6pt}  \\ 
		     \tilde{A}_2-\tilde{A}_1\frac{a_{20}}{a_{10}} \\
			     :                                    \\
		     \tilde{A}_r-\tilde{A}_1\frac{a_{r0}}{a_{10}}
    \end{array} \right)  \label{d8}
  \end{equation}
  \begin{equation}
    a_{00}  \rightarrow  - \tilde{A}_1 \left(\frac{a_{00}}{a_{10}}\right) \label{d9}
  \end{equation}
and add (\ref{d6}) to the list of substitutions. Updating components $2 \ldots r$ in (\ref{d8})
is done purely by multiplications and additions, only updating the first component takes 
differentiations when factoring out $D$.

A more conventional algorithm solving underdetermined linear ODEs tries to lower the order
of the ODE by performing Euclids algorithm until the ODE is algebraic for one function
and thus can be solved algebraically. In the following section we formulate such an
Euclidean step to be able to compare it with the new algorithm above.
  
\section{The Euclid version in vector notation}  \label{euclid}
For applying the `right' Euclid algorithm the ODE is given as well in a form with
$D$ completely factored out:
\begin{eqnarray}
	 0   & = & \sum_i^r A_i f_i \, + \, a_{00}(x)  \label{f1} \\
        A_i  & = & D^{n_i} a_{in_i}(x)+\ldots+Da_{i1}(x)+a_{i0}(x) .  \label{f2} 
\end{eqnarray}
One iteration step is performed by
\begin{itemize}
\item choosing two functions $f_i, f_j$ (w.l.o.g.\ $n_i \geq n_j$) and
      introducing a new function $f_{r+1}$ through
  \begin{equation}
    f_j = f_{r+1} - D^{n_i-n_j}\left(\frac{a_{in_i}}{a_{jn_j}}f_i\right) \label{f3} 
  \end{equation}
\item and performing this substitution in the ODE (\ref{f1}).
\end{itemize}
The substitution is chosen such that the differential order of $f_i$ is lowered 
by at least one. This iteration process stops like in the new algorithm when the first function appears
purely algebraically or when the ODE involves only one function.
Therefore, in order to minimize the number of steps one would
choose $f_j$ as one of the lowest order functions and $f_i$ as one of the
lowest order functions from the remaining ones. (From all these choices it
is recommendable to choose a pair so that multiplication is minimized
and $A_i, A_j$ have as few as possible terms.)
For $j=1, i=2$ the vector notation gives 
  \begin{equation}
    \left(\begin{array}{c} A_1 \vspace{6pt} \\ A_2 \vspace{6pt} \\ A_3 \\ : \\ A_r \end{array}\right) \rightarrow 
    \left(\begin{array}{c} A_{r+1} \vspace{6pt} \\ A_2 \vspace{6pt} \\ A_3 \\ : \\ A_r \end{array}\right) = 
    \left(\begin{array}{c} A_1 \vspace{6pt} \\ A_2-A_1D^{n_2-n_1}\frac{a_{2n_2}}{a_{1n_1}} \vspace{3pt} \\
		     A_3 \\  :  \\ A_r
    \end{array} \right)  \label{f4} 
  \end{equation}
  \begin{equation}
    a_{00} \rightarrow a_{00} . \label{f5} 
  \end{equation}
The update of the $2^{\rm nd}$ component in (\ref{f4}) requires 
differentiations when $D$ is factored out as far as possible. 

\section{Relations between both algorithms}  \label{compare}
\subsection{Differences} \label{differences}
In this section we want to justify our claim that both algorithms
differ significantly from each other and that they are not merely
variations of one and the same procedure.

Both algorithms differ conceptually in that Euclids algorithm pairs
two differential operators to get a new differential operator of lower
order whereas in the new algorithm the relation that defines a new
function becomes the ODE and the old ODE is used for substituting one
of the `old' functions.

Another difference is that Euclids algorithm lowers in one step the
order of only one differential operator whereas the new algorithm
lowers the order of all but one differential operators. Both types of
steps are performed at about the same cost as for the new method only
the update of the first component in (\ref{d8}) is potentially size
increasing and for Euclids method it is only the update of the
second component of (\ref{f4}). Using Euclids algorithm one can of
course take one of the lowest order differential operators and
decrease the order of all others but that would take $r-1$
computations, each potentially size increasing.  This difference does
usually not affect the number of necessary iterations (until one
function occurs purely algebraically), because Euclids method can
operate just on two of the lowest order operators and ignore all
others.  So, by applying Euclids method to lower the order of the two
lowest order operators $A_i$ with each other, all other higher order
operators keep their high order. Moreover, when the algorithm stops
because the final equation contains one function algebraically, then
back substitutions start (part B of the algorithm) 
which even increase the highest orders.

Differently with the new method where the differential order of all
functions but one gets lowered in one step. Thus the obtained parametric 
solution is typically of lower order than the solution obtained
by Euclids algorithm if this is executed by pairing only lowest order derivatives.

For example, the size of the parametric solution of the ODE
\begin{equation}
0 = x^3f_{ax} + (x-1)g_{bx} + h_{5x}, \ \ \ \mbox{for}\ f,g,h\ \mbox{of}\ x \label{g1}
\end{equation}
and the differential order of parametric functions in it 
depend strongly on the differential orders $a,b$ of $f$ and $g$ in the ODE
and the method that is used. 
As $a,b$ increase from $a=b=1$ to $a=b=3$ one can see the following trends:


\begin{itemize}
\item In Euclids solution the differential order of parametric functions
      increases from 5 to 7 whereas in the solution of the new algorithm
      the order decreases from 4 to 3 in the expression for $g$, from 4 to 2 in 
      the expression for $f$ and only increases from 0 to 1 in the expression for $h$.
\item The size of expressions in Euclids solution steadily increases and
      in the solution of the new method it decreases. 

      For $a=b=1$ the solutions have comparable size: \\
      - When explicit $x$-dependent factors are not absorbed 
      (for absorbing factors see section \ref{absorb})
      then $h$ is parametric and the rational expressions for $f,g$
      have the form (5 terms)/(2 terms) in Euclids solution and in the solution
      of the new method $f=$ (4 terms)/(2 terms), $g=$ (3 terms)/(2 terms). 

      For $a=b=3$:\\
      - When factors are not absorbed then in Euclids solution $h$ is parametric,
      $f=$ (85 terms)/(15 terms), $g=$ 13 terms and in the solution
      of the new algorithm we have $f=$ 1 term, $g=$ 7 terms, $h=$ 4 terms. \\
      - When factors are absorbed then Euclids solution gives  
      $f=$ 27 terms, $g=$ 16 terms and in the solution
      of the new algorithms we have $f=$ 1 term, $g=$ 7 terms, $h=$ 4 terms. 
\end{itemize}

%
%


Not only the solutions of both algorithms may differ significantly but
also the number of steps to reach them.
Starting, for example, with
\[0 = f' + a(x) h^{(n)}=0 \]
to be solved for $f=f(x), h=h(x)$ 
this takes only one iteration step with the new algorithm 
giving, for example for $n=5$, a solution of the form 
$f=$ (27 terms)/(1 term), $h=$ (1 term)/(1 term), 
whereas it takes $n$ steps for Euclids algorithm giving, for example for $n=5$,
a more spacious solution of the form $f=$ (40 terms)/(1 term), $h=$ (2 terms)/(1 term).

If the ODE is inhomogeneous then another difference between both 
algorithms becomes apparent. With non-zero inhomogeneity $a_{00}$ 
in the ODE (\ref{d1}) the transformation (\ref{d6}) becomes 
inhomogeneous and changes $a_{00}$ in (\ref{d9}) 
whereas Euclids algorithm does not change $a_{00}$ 
in the homogeneous transformation (\ref{f3}). 

Two ODEs that highlight the duality between both algorithms are
\begin{eqnarray}
0 & = & f' + f + g' + (ah)^{(20)}  \label{ex2}  \\
0 & = & f' + f + g' + a(h)^{(20)}  \label{ex3}
\end{eqnarray}
where $a=a(x)$ is a given function of $x$ and the ODE 
has to be solved for $f,g,h$.
As shown in table 1 for the ODE (\ref{ex2}) the new algorithm is to be preferred
whereas for ODE (\ref{ex3}) Euclids algorithm gives a shorter solution.
\begin{center}
\begin{tabular} {|l|ll|ll|} \hline
algorithm     & \multicolumn{2}{c|}{equation (\ref{ex2})}        
              & \multicolumn{2}{c|}{equation (\ref{ex3})}                \\ \hline
new           & $f$=1 term   & $g$=22 terms & $f$=2 terms & $g$=23 terms \\ \hline
Euclid        & $f$=22 terms & $g$=23 terms & $f$=2 terms & $g$=3 terms  \\ \hline
\end{tabular} \vspace*{6pt} \\
Table 1. The size of solutions given by both algorithms (both executed without 
absorbing of factors).
\end{center}

An explanation of the above behaviour comes from the fact that
Euclids method is better suited when factors
$\frac{a_{in_i}}{a_{jn_j}}$ are small, i.e.\ when the coefficients of
the leading derivatives in the representation (\ref{d1}) -
(\ref{d3}) effectively cancel each other in a quotient, and less
suited when these factors are large prime expressions in
$x$. Differently, the new method works best if factors
$\frac{a_{i0}}{a_{j0}}$ are small, i.e.\ when the coefficients of the
algebraic terms 
effectively cancel each other, and less suited when these factors are
large prime expressions in $x$.

Although both ODEs look similar, they are rather different as
$(ah)^{(20)}$ has many terms if the product rule of differentiation is
fully applied and $a(h)^{(20)}$ has many terms if $\frac{d}{dx}$ is
factored out as far as possible. Thus both algorithms differ in their
suitability for both ODEs.

The above computations and any other tests can easily be performed
online (see \cite{Wuode}), where access to the computer algebra system
REDUCE and to the procedures {\tt uode} and {\tt print\_uode\_solution}
is provided.


\subsection{Hybridization} \label{hybrid}
The advantage of having these two, in some sense, complementary
algorithms lies in the fact that both operate on the same data
structure, i.e.\ both their input and output consists of an ODE in the
form (\ref{d1}) - (\ref{d3}) and both add one substitution to a list
of substitutions of functions in terms of newly introduced
functions. Consequently, both algorithms are interchangeable, i.e.\ one
step performed with one algorithm could be followed by another step
performed with the other algorithm, in order to minimize the size of
$x$-dependent factors and thus to minimize growth. A different
strategy could be to perform each step with both algorithms in parallel, 
to compare the size of the ODE and/or size of derived
substitution that both methods give and to choose which one to adopt
for this step. This strategy does at most
double the amount of computation but more likely will lead to savings
due to working with smaller expressions.

Any hybrid algorithm performing a mixture of Euclid steps and new algorithm steps
is still finite because in any such step the sum of differential orders
of all functions is decreasing.

\subsection{Absorbing factors} \label{absorb}
The following efficiency improving measures work for both methods.
A representation (\ref{d1}) - (\ref{d3}) (and identically (\ref{f1}) - (\ref{f2}))
of the ODE where $D$ is maximally factored out has the advantage
that any change of functions $f_i \rightarrow h(x)\bar{f}_i$ does require only
multiplications and is done very easily in both algorithms. 
This freedom of efficiently multiplying functions with $x$-dependent factors can be
used for different purposes.

If in a newly generated ODE all coefficients of a function $f_i$ have a non-trivial GCD 
$\hat{c}(x):=GCD(a_{in_i},\ldots,a_{i0}) \neq 1$ then this can be absorbed 
into a new function $f_{r+1}$ and the ODE be simplified by performing the substitution 
\begin{equation}
 f_i = f_{r+1}/\hat{c}  \label{f6}
\end{equation}
and adding it to the accumulating list of substitutions. Such non-trivial GCDs
occur relatively frequently as the GCD is taken only from the few coefficients $a_{ij}$
of any single one operator $A_i$ that changed in an iteration step. 

A different purpose of introducing new functions multiplied with an explicit
$x$-depending factor is to avoid a denominator (den). This would arise in the first component of
(\ref{d8}) and it can be prevented by introducing another new function $f_{r+2}$ and 
by performing the substitution
\begin{equation}
 f_{r+1} = a_{10}^{\;\;2} f_{r+2} .  \label{f7}
\end{equation}
Similarly, in the other components of (\ref{d8}) one can scale 
\begin{equation}
 f_i = \mbox{den}\left(\frac{a_{i0}}{a_{10}}\right) f_{r+i}.   \label{f8}
\end{equation}

In Euclids algorithm the appearance of denominators in the second component of (\ref{f4})
can be prevented by introducing the scaling
\begin{equation}
 f_2 = \mbox{den}\left(\frac{a_{2n_2}}{a_{1n_1}}\right) f_{r+1}.   \label{f9}
\end{equation}

Another situation where a dominator occurs is at the end of both algorithms
when one function occurs purely algebraically 
\begin{equation}
 0 = a_{i0}f_i + \sum_{j \neq i} A_j f_j \, + \, a_{00}(x).  \label{f10} 
\end{equation}
To avoid the denominator $a_{i0}$ one can scale all $f_j$ depending on their
differential order and on their coefficients in $A_j$.
Avoiding this denominator is especially helpful, as $f_i = \ldots$ is the 
last substitution and thus substituted successively in all previous substitutions 
leading easily to a substantial growth of denominators. 
By avoiding denominators in the above way  both methods produce 
denominator free solutions if the ODE is homogeneous.

\subsection{Embedded ODEs}  \label{factor}
The case that an underdetermined ODE factorizes, i.e. that 
the differential operators $A_i$ in (\ref{d1}) have a non-trivial GCD, 
or in other words, that the ODE can be written in form of two nested ODEs
$0 = \Omega(x,\omega(x,f_i))$ is discovered by both algorithms. The inner ODE
$0=\omega$ is only determined up to a linear change $\omega=\alpha(x)\hat{\omega}$
and both algorithms will usually find $\Omega$'s that differ by some $\alpha(x)$.
A slight advantage of the new algorithm is that the $D^0$ part of the ODE
is used within the algorithm, thus it is automatically recognized if this vanishes,
i.e.\ if the ODE is exact (up to an inhomogeneity).

\section{ODE-systems}  \label{systems}
The introduced methods of solving a single underdetermined ODE
(or converting it to an ODE for a single function) can be used
to convert an ODE system into an equivalent set of fully decoupled ODEs,
each for a single function.

After treating any one equation of the original ODE system the
original functions in it are expressed in terms of fewer functions
which are either all free or at most one has to satisfy a single ODE.
All functions can be replaced in the remaining ODEs. This can go on as
long as we have equations of at least two functions. When the
procedure stops we have solved the system or got ODEs, each containing
only one function. If more than one ODE contain only one and the same
single function, then these ODEs form an over-determined subsystem
which can be treated by a Gr\"{o}bner bases computation and result
either in the explicit solution for this function or a single ODE for
this function of an order not higher than the lowest order of the ODEs
for this function. This whole procedure terminates with either the
explicit parametric solution of the original system or a decoupled set
of ODEs, each ODE for a single function.

%
%

\section{An application} \label{application}
A class of applications where underdetermined linear ODEs occur
frequently is the classification of hyperbolic evolutionary PDE
systems. The aim of such an investigation is to find integrable systems
of PDEs by determining those systems which have a higher order symmetry 
(see below). More information about the mathematical background is given 
in \cite{AncoWolf04} where a classification of 
hyperbolic {\em vector} PDEs is discussed. 

Let us look at two hyperbolic scalar PDEs for functions $u(x,t,\tau),
v(x,t,\tau)$ (where $x,t$ are the usual independent variables and
$\tau$ is a symmetry parameter).
The following ansatz for the system and symmetry is generated based
on homogeneity considerations. 
By requiring the same homogeneity weights as the potential 
nonlinear Schr\"{o}dinger equation we get for the system the ansatz
\begin{eqnarray} 
 u_{tx}&=& 
a_{10}u_{3x} + 
a_{6}u_xv_x^2u^2 +
a_{8}u_{2x}u_xv + 
a_{9}u_{2x}v_xu + 
a_{1}u_x^2v_x + 
a_{3}u_x^3v^2 + 
a_{7}u_xv_{2x}u 
           \nonumber \\ 
 v_{tx}&=& 
a_{19}v_{3x} + 
a_{15}u_x^2v_xv^2  + 
a_{18}v_{2x}v_xu + 
a_{17}u_xv_{2x}v + 
a_{13}u_xv_x^2 + 
a_{16}v_x^3u^2 . 
            \label{he2}
\end{eqnarray}
%
Assuming the same differential order for the symmetry
we get the ansatz 
\begin{equation} 
 u_\tau = b_{10}u_{3x} + \ldots , \ \ \ \  
 v_\tau = b_{19}v_{3x} + \ldots , \label{hs}
\end{equation}
with right hand sides identical to those of (\ref{he2}),
only with coefficients $b_i$ instead of $a_i$. All coefficients $a_i, b_i$ are
undetermined functions of the product $uv$.\footnote{For hyperbolic systems
the homogeneity weights include negative values, for example here weight($u$)=1,
weight($v$)=-1, so that $uv$ has weight zero and therefore we have no limitation on the
degree of powers of $uv$ and thus all unknown coefficients are arbitrary functions
of $uv$.}

The relations (\ref{hs}) are  
considered to be a symmetry of the system (\ref{he2}) if the symmetry conditions
\begin{equation}
   \partial_\tau (u_{tx}) - \partial_t\partial_x (u_\tau) = 0 , \ \ \ \ \ 
   \partial_\tau (v_{tx}) - \partial_t\partial_x (v_\tau) = 0 \label{scon}
\end{equation}
are fulfilled identically in $u,v$ and derivatives of $u$ and $v$ 
both modulo substitutions based on (\ref{he2}) and (\ref{hs}).
In performing the differentiations in (\ref{scon}), doing repeatedly
substitutions (\ref{he2}), (\ref{hs}) and finally setting all
coefficients of different products of powers of derivatives of $u,v$
individually to zero gives 27 ODEs with a total of 1334 terms for 26
functions $a_i,b_j$ of $z:=uv$.  The length of equations ranges from 2
to 266 terms.  In the course of solving this system the program {\sc
Crack} performs integrations, substitutions, splittings (i.e.\
separations when $z$ occurs only explicitly) and a number of
case distinctions. In one of the sub cases the resulting conditions can be
integrated successively up to the underdetermined linear ODE
\begin{equation}
0 = 3b_{13}^{\ \ '}z - 6b_{15}^{\ \ '}z^2  - 2b_{17}^{\ \ ''}z^2  
      + b_{17}^{\ \ '}z - 6b_{15}z + 2b_{17}. \label{ex1a} 
\end{equation}
{\em Step 1:} To start, partial integration gives
\begin{equation}
0=(3b_{13}z-6b_{15}z^2-2b_{17}^{\ \ '}z^2+5b_{17}z)^{'}-3b_{13}+6b_{15}z-3b_{17}.  \label{ex1c}
\end{equation}
which can be written as
\begin{equation}
0=c_1^{\ '}-3b_{13}+6b_{15}z-3b_{17}  \label{ex1e}
\end{equation}
by introducing $c_1(z)$ through 
\begin{equation}
c_1=3b_{13}z-6b_{15}z^2-2b_{17}^{\ \ '}z^2+5b_{17}z. \label{ex1d}
\end{equation}
From the functions that occur only algebraically in (\ref{ex1e}) 
(i.e.\ $b_{13},b_{15},b_{17}$)
the ones that have lowest derivatives in (\ref{ex1a}) are $b_{13},b_{15}$. 
Eliminating one of them from (\ref{ex1e}), say $b_{13}$
and substituting it in (\ref{ex1d}) gives
\begin{equation}
0=2b_{17}^{\ \ '}z^2-2b_{17}z-c_1^{\ '}z+c_1  \label{ex1g}
\end{equation}
which is not algebraic in any function yet, but already of first order, 
so one more step has to be
performed. 

{\em Step 2:} Partial integration of (\ref{ex1g}) and introduction of 
\begin{equation}
c_2=2b_{17}z^2  - c_1z    \label{ex1h}
\end{equation}
results in 
\begin{equation}
0= c_2^{\ '} - 6b_{17}z + 2c_1    \label{ex1i}
\end{equation}
which allows to solve for $c_1$ and replace it in (\ref{ex1h}) giving 
\begin{equation}
0= \frac{1}{2}c_2^{\ '}z - c_2 - b_{17}z^2.     \label{ex1ii}
\end{equation}
This condition is purely algebraic for one function, $b_{17}$, and 
therefore the algorithm stops. 

{\em Cleanup:} We need explicit solutions of (\ref{ex1a}), so 
what remains to be done are back substitutions: (\ref{ex1ii}) provides
\begin{equation}
b_{17}=\frac{1}{2z}c_2^{\ '} - \frac{1}{z^2}c_2.    \label{ex1j}
\end{equation}
The second substitution expressing $b_{13}$ in terms of the parametric
function $c_2$ is obtained after backward substituting 
$c_1$ from (\ref{ex1i}) into (\ref{ex1e}) and 
$b_{17}$ from (\ref{ex1j}) in (\ref{ex1e}) to get 
the explicit solution consisting of (\ref{ex1j}) and 
\begin{equation}
b_{13}=\frac{1}{3}c_2^{\ ''}-\frac{3}{2z}c_2^{\ '}+2b_{15}z+\frac{2}{z^2}c_2  \label{ex1k}
\end{equation}
involving the free function $c_2(z)$.

Another underdetermined equation resulting in this integrability problem is
\begin{equation}
0 = 3b_1^{\ '}z - 9b_3^{\ '}z^2  + 3b_6^{\ '}z^2  - 4b_8^{\ ''}z^2  
    - 12b_8^{\ '}z - 12b_3z + 6b_6z - 3b_8
\label{ex1b} 
\end{equation}
which has the solution
\begin{eqnarray*}
b_1 &=& ( - 8b_8^{\ '}z^2 + 3c_3^{\ '}z + 6b_6z^2 - 5b_8z - 2c_3)/(3z)  \label{ex1m} \\
b_3 &=& ( - 4b_8^{\ '}z^2 + c_3^{\ '}z + 3b_6z^2 - 3b_8z - c_3)/(3z^2).   \label{ex1n} 
\end{eqnarray*}

We finally obtain as a system with higher order symmetries:

%
%
%

\begin{eqnarray*}
u_{tx} & = & \frac{a_1}{2uv^3}\left( (uv)_x^{\ 3}v-uv(uv)_x^{\ 2}v_x \right) \\
       &   & \\
v_{tx} & = & \frac{a_{13}}{2uv}(uv)_x^{\ 2}v_x .
\end{eqnarray*}

\section{Appendix} \label{appendix}
The following example illustrates the comments made in the last paragraphs
of section \ref{comments}. It shows two possible representations
of a solution to an underdetermined ODE. For the equation 
\[ (x-1)^3f_1^{(5)}+3f_1^{(3)} + xf_1'' + (1-x^2)f_1' + f_1 - (x-2)(x-3)f_2'' - xf_2' = 0 \]
the generated list $L$ of substitutions is given through

\small
\begin{eqnarray*}
f_6&=&(7289/9 f_8' x^6 - 39400/3 f_8' x^5 + 725945/9 f_8' x^4 - 
    691667/3 f_8' x^3 + 2665016/9 f_8' x^2 \\
   & &- 463541/3 f_8' x + 40582 f_8' - 82543/9 f_8 x^5 + 393212/3 f_8 x^4 - 1997777/3 f_8 x^3 \\
   & &+ 1404610 f_8 x^2 - 9064364/9 f_8 x + 863648/3 f_8)/(x^12 - 27 x^11 + 339 x^10
    - 2551 x^9 \\
   & &+ 12566 x^8 - 43294 x^7 + 111667 x^6 - 221121 x^5 + 332143 x^4
    - 447477 x^3 + 625912 x^2 \\
   & &- 422746 x + 117948) \\
f_7&=&( - 197/419 f_8' - 2227/6704 f_6 x^6 + 40655/6704 f_6 x^5 - 282747/
    6704 f_6 x^4 \\
   & &+ 937545/6704 f_6 x^3 - 737203/3352 f_6 x^2 + 236215/1676 f_6 x 
    - 15636/419 f_6)/ \\
   & &(x^5 - 5988/419 x^4 + 30423/419 x^3 - 64170/419 x^2 + 46012
    /419 x - 13152/419) \\
f_1&=&( - 288/197 f_7' - 450/197 f_6)/(x^5 - 2988/197 x^4 + 15683/197 x^3 - 
    32571/197 x^2 \\
   & &+ 18956/197 x - 6093/197) \\
f_5&=& - 9/16 f_6' + 37/32 f_1 x^4 - 415/32 f_1 x^3 + 347/8 f_1 x^2 - 815/32 f_1 x
    + 261/32 f_1 \\
f_4&=& - 4/9 f_5' - 17/9 f_1 x^3 + 124/9 f_1 x^2 - 176/9 f_1 x + 11 f_1 \\
f_3&=& - 1/4 f_4' + 5/4 f_1 x^2 - 17/4 f_1 x - 11/4 f_1 \\
f_2&=&f_3' + 2 f_1 x + f_1
\end{eqnarray*}
\normalsize
which is a much shorter representation of the solution than the
10 page explicit form which results from substituting $f_6,f_7,f_1,f_5,f_4,f_3$ in
this order into each other and takes the form $f_1=$(42 terms)/(25
terms), $f_2=$(304 terms)/(61 terms) involving up to 29-digit integers.
The list of 7 substitutions is not only shorter than the list of 2 substitutions
for $f_1,f_2$, it also is much faster to derive and more useful if a differential
expression is to be simplified modulo the solution of the above ODE by
substituting $f_2,f_3,f_4,f_5,f_1,f_7,f_6$ in this order.

The difference in size of both solution representations can be
arbitrarily amplified by having an input ODE of higher order and higher
degree polynomials as coefficients.

\section{Summary}  \label{summary}
We present an algorithm for the solution of underdetermined linear ODE
that is compatible but structurally different from the (right)
Euclidean algorithm.  Because both algorithms operate on the same data
structure and because both complement each other in the sense that
each one is most efficient for ODEs of different form, the combination
of both algorithms is superior to each individual one.

\section*{Acknowledgements}
I would like to thank Sergey Tsarev for many comments. 
Daniel Robertz is thanked for comparative runs with 
the OreModules package.


\begin{thebibliography}{99}
\bibitem{CQR05}Chyzak, F., Quadrat, A., Robertz, D. (2005). "Effective 
 algorithms for parametrizing linear control systems over Ore algebras", 
 Rapport de Recherche INRIA ,  Applicable Algebra in Engineering, 
 Communications and Computing {\bf 16}, no 5, 319-376. 
\bibitem{PQ04}Pommaret, J.-F., Quadrat, A. (2004). "A differential operator 
 approach to multidimensional optimal control", International Journal of 
 Control {\bf 77}, 821-836.
\bibitem{FO98}Fr\"{o}hler, S., Oberst, U. (1998). "Continuous time-varying 
 linear systems", Systems \& Control Letters {\bf 35}, 97-110.
\bibitem{CQRwww}
 Chyzak, F., Quadrat, A., Robertz, D. 
 OreModules project \\ \verb+http://wwwb.math.rwth-aachen.de/OreModules+
\bibitem{CQR07}
 Chyzak, F., Quadrat, A., Robertz, D. "OreModules: A symbolic package for 
 the study of multidimensional linear systems" in: J. Chiasson, J.-J. Loiseau,
 "Applications of Time-Delay Systems", Springer, to appear.
\bibitem{AncoWolf04}Anco, S.\ and Wolf, T. (2005). "Some symmetry 
 classifications of hyperbolic vector evolution equations", JNMP, Volume 12, 
 Supplement 1, p 13-31 (also nlin.SI/0412015).
\bibitem{Wuode}
 Wolf, T. (2007). Online demo for solving underdetermined ODEs. \\
 {\tt http://lie.math.brocku.ca/crack/uode}
\bibitem{crack} Wolf, T.: Applications of {\sc Crack} in the Classification of
 Integrable Systems, CRM Proceedings and Lecture Notes,
 vol 37 (2004) pp. 283-300. (arXiv nlin.SI/0301032) and online under
 {\tt http://lie.math.brocku.ca/crack/demo}
\end{thebibliography}
\end{document}